\documentclass[aps,pre,reprint,twocolumn,showpacs,superscriptaddress,groupaddress,showkeys,floatfix,byrevtex]{revtex4-2}
\usepackage{amsmath}
\usepackage{amsfonts,bm}
\usepackage{amssymb}
\usepackage[dvipsnames]{xcolor}
\usepackage{graphicx,color,subeqnarray}
\usepackage{dcolumn}
\usepackage{bm}
\usepackage{soul}
\usepackage{upgreek}

\usepackage{srcltx}
\makeatletter
\DeclareFontFamily{OMX}{MnSymbolE}{}
\DeclareSymbolFont{MnLargeSymbols}{OMX}{MnSymbolE}{m}{n}
\SetSymbolFont{MnLargeSymbols}{bold}{OMX}{MnSymbolE}{b}{n}
\DeclareFontShape{OMX}{MnSymbolE}{m}{n}{
    <-6>  MnSymbolE5
   <6-7>  MnSymbolE6
   <7-8>  MnSymbolE7
   <8-9>  MnSymbolE8
   <9-10> MnSymbolE9
  <10-12> MnSymbolE10
  <12->   MnSymbolE12
}{}
\DeclareFontShape{OMX}{MnSymbolE}{b}{n}{
    <-6>  MnSymbolE-Bold5
   <6-7>  MnSymbolE-Bold6
   <7-8>  MnSymbolE-Bold7
   <8-9>  MnSymbolE-Bold8
   <9-10> MnSymbolE-Bold9
  <10-12> MnSymbolE-Bold10
  <12->   MnSymbolE-Bold12
}{}

\let\llangle\@undefined
\let\rrangle\@undefined
\DeclareMathDelimiter{\llangle}{\mathopen}
{MnLargeSymbols}{'164}{MnLargeSymbols}{'164}
\DeclareMathDelimiter{\rrangle}{\mathclose} {MnLargeSymbols}{'171}{MnLargeSymbols}{'171}
\makeatother

\begin{document}

\title{Nonequilibrium dynamical structure of a dilute suspension of active particles in a viscoelastic fluid}

\author{Juan Ruben \surname{Gomez-Solano}}
\email{r\_gomez@fisica.unam.mx}
\affiliation{Instituto de F\'isica, Universidad Nacional Aut\'onoma de M\'exico, Ciudad de M\'exico, C\'odigo Postal 04510, Mexico,}

\author{Rosal\'io F. \surname{Rodr\'iguez}}
\affiliation{Instituto de F\'isica, Universidad Nacional Aut\'onoma de M\'exico, Ciudad de M\'exico, C\'odigo Postal 04510, Mexico,}
\affiliation{FENOMEC, Universidad Nacional Aut\'onoma de M\'exico, Apdo. Postal 20-726, 01000, Ciudad de M\'exico, Mexico,}

\author{Elizabeth \surname{Salinas-Rodr\'iguez}}
\affiliation{Departamento I. P. H., Universidad Aut\'onoma Metropolitana, Iztapalapa, Apdo. Postal 55–534, 09340 Ciudad de M\'exico, Mexico.}

\date{\today}

\begin{abstract}
In this work, we investigate the dynamics of the number density fluctuations of a dilute suspension of active particles in a linear viscoelastic fluid. We propose a model for the frequency-dependent diffusion coefficient of the active particles, which captures the effect of rotational diffusion on the persistence of their self-propelled motion and the viscoelasticity of the medium. Using fluctuating hydrodynamics, the linearized equations for the active suspension are derived, from which we calculate its dynamic structure factor and the corresponding intermediate scattering function. For a Maxwell-type rheological model, we find an intricate dependence of these functions on the parameters that characterize the viscoelasticity of the solvent and the activity of the particles, which can significantly deviate from those of an inert suspension of passive particles and of an active suspension in a Newtonian solvent. In particular, in some regions of the parameter space we uncover the emergence of oscillations in the intermediate scattering function at certain wave numbers, which represent the hallmark of the non-equilibrium particle activity in the dynamical structure of the suspension and also encode the viscoelastic properties of the medium.
\end{abstract}

\maketitle

\section{Introduction}

Active matter, which has recently emerged as a prominent research field of physical sciences \cite{gompper2020}, focuses on systems that are made up of autonomous entities capable of consuming energy from their environment in order to convert it into directed motion \cite{ramaswamy2010}. This definition encompasses a broad diversity of non-equilibrium systems ranging from locomotive macroscopic animals to mesoscopic biomolecular motors \cite{haenggi2009,elgeti2015,bechinger2016,grosswasser2018}. Of special interest is the study of active soft materials, such as the cytoplasm of eukaryotic cells \cite{juelicher2007,prost2015,koslover2017,mogilner2018}, seminal fluids \cite{riedel2005,yang2008,tung2017,ishimoto2018}, bacterial colonies \cite{dombrowski2004,dellarciprete2018,beer2019,ramos2021} and aqueous suspensions of synthetic active colloids \cite{howse2007,palacci2010,jian2010,qian2013,gomezsolano2017,narinder2019,gomez_solano2020}, which exhibit complex fluid-like structural, dynamical and rheological properties that significantly differ from those of inert fluids \cite{marchetti2013}. For instance, unlike passive colloids dispersed in a viscous solvent, which increase the total viscosity of the resulting suspension \cite{verberg1997}, the addition of swimming bacteria can reduce it \cite{cates2008,sokolov2009}. Moreover, at higher densities, remarkably complex effects can emerge, ranging from active superfluid behavior \cite{lopez2015} to turbulent flows of concentrated bacterial suspensions at low Reynolds number \cite{dunkel2013}. Therefore, there is a growing interest in understanding collective effects in active soft matter, as they open up new avenues for designing novel soft materials with specific functionalities \cite{needleman2017,mallory2018}. This requires the use of experimental techniques that, in addition to conventional single-particle tracking methods \cite{ebbens2011,dupont2013,liu2016}, are able to provide ensemble information of dynamical and structural properties of active-particle suspensions. For instance, dynamical light scattering (DLS) \cite{berne1976}, differential dynamic microscopy (DDM) \cite{cerbino2008}, and super-heterodyne laser-Doppler-velocimetry (SH-LDV) \cite{botin2017}, have been successfully applied to the characterization of aqueous suspensions of Janus colloidal particles \cite{lee2014,wittmeier2015,kurzthaler2018,sachs2021}, motile bacteria \cite{wilson2011,lu2012,martinez2012,linek2016}, and cytoskeletal filamentous actin \cite{drechsler2017}. Furthermore, analytical expressions for the intermediate scattering function, a quantity directly accessible by DDM, have been derived for stochastic models of active matter, such as run-and-tumble particles \cite{martens2012}, active Brownian particles \cite{kurzthaler2016,kurzthaler2017}, and oscillatory breaststroke microswimmers \cite{croze2019}. Although such ensemble techniques have proved to be valuable for investigating specific dynamical details of active particles in viscous solvents under homogeneous conditions, other situations of practical interests remain largely unexplored. In particular, the effect of viscoelasticity of the solvent on the bulk properties of the active suspension due to the presence of suspended macromolecules could be of great significance for many fluids of biological and technological importance \cite{denn2004,phan2012,zhou2020,wu2020,li2021}.

The main purpose of this work is to get insights into the effects produced by the hydrodynamic fluctuations of a viscoelastic fluid on a suspension of active particles. The presence of time memory kernels due to viscoelasticity induces thermally excited retarded fluctuations, a situation that does not only occur in viscoelastic fluids, but also in glassy materials \cite{zaccone2020}, and is of increasing importance for complex biological media \cite{vitali2018}. Although these effects have been theoretically studied mainly for Brownian motion and diffusive systems \cite{wang1999,metzler2000,rodriguez2013,wang2020,rodriguez2021}, much less is known of their implications on the fluctuations and transport properties of active hydrodynamic systems.
The dynamics of the solvent fluctuations around its quiescent state are described by using fluctuating hydrodynamics  \cite{fox1978,ortiz2006}. On the other hand, to describe the motion of the active particles through the viscoelastic fluid, we propose a stochastic model for their diffusion coefficient. This
model, which is motivated by experimental observations, captures qualitatively the effect of rotational diffusion in the persistence of self-propulsion with a characteristic propulsion speed. It also depends explicitly on the type of viscoelasticity of the medium. The general expressions for the frequency-dependent diffusion coefficient are then written for Newtonian solvents as well as for Maxwell-type viscoelastic fluids. This approach leads to analytical expressions for the  dynamic structure factor of the suspensions in terms of the parameters that characterize the viscoelasticy of the medium as well as the particle activity. Using such expressions, we numerically compute the intermediate scattering function and analyze the distinct behaviors of the dynamical structure of the active suspension that can emerge in its parameter space.

The plan of the paper is as follows. In Sec. \ref{sect:formul} we introduce the relevant features of the
model for a dilute  suspension of active particles in a viscoelastic fluid. The hydrodynamic equations for the particles and solvent are formulated, and the non-equilibrium steady state (NESS) to be considered is defined. In Sec. \ref{sect:fluct} the linearized fluctuating hydrodynamic equations are
formulated, which are then solved in Fourier domain for the fluctuations of the number density of active particles and the velocity fluctuations of the solvent. In Sec. \ref{sect:diff} a model for the frequency-dependent diffusion coefficient of the active particles in the suspension is proposed, which encodes both the viscoelasticity of the fluid and the properties of the active particles. From it, a general formal expression for the dynamic structure factor is obtained in Sec. \ref{sect:dsf}, which is then applied to investigate in detail the intermediate scattering function of dilute suspensions of active particles in a Newtonian solvent and in a Maxwell-type viscoelastic fluid. Finally, in Sec. \ref{sect:summ}, we summarize the main results of our work and make some further physical remarks.

\section{Formulation of the model}\label{sect:formul}

\begin{figure*}\centering
 \includegraphics[width=0.8\textwidth]{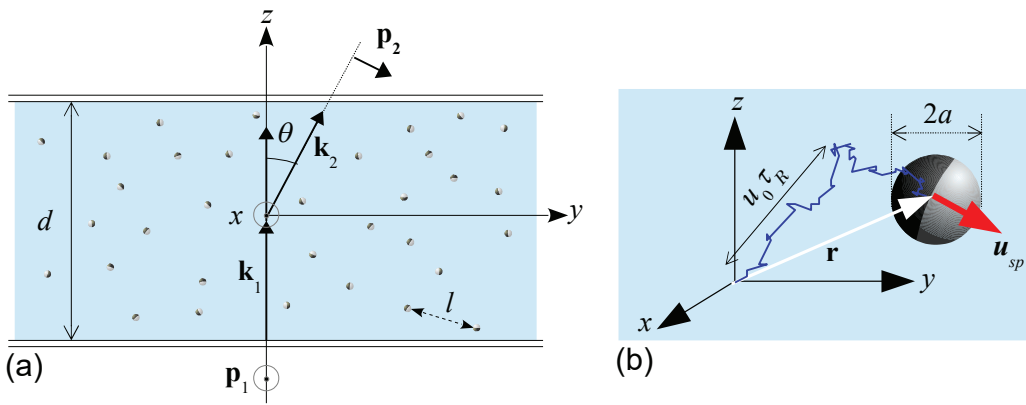}
 \caption{(a) Viscoelastic fluid layer between two parallel plates with separation $d \gg a$, where $a$ is the radius of the embedded active particles, which are separated on average by a distance $l$ such that $a \ll l \ll d$. Some quantities describing a light scattering process have been also included: The $xz$ plane is the scattering plane. The angle $\theta$ between the wave vector of the incident light, $\mathbf{k}_1$, and the one of the outgoing beams, $\mathbf{k}_2$, is the scattering angle. The wave vector of the scattered light, $\mathbf{k}_1 - \mathbf{k}_2$, is related to $\theta$ by $k = |\mathbf{k}| = 4 \pi n \sin (\theta/2)/\lambda$, where $\lambda$ is the wavelength of the incident light beam in vacuum, and $n$ is the refractive index of the scattering medium. $\mathbf{p}_1$ and $\mathbf{p}_2$ denote the polarizations of the incident and scattered beams, respectively. (b) Active particle moving in the viscoelastic fluid with propulsion velocity $\mathbf{u}_{sp}$ that evolves according to the Ornstein-Uhlenbeck process described by Eq. (\ref{eq:OU}). On the time scale $\tau_R$, the particle exhibits persistent motion with a persistence length $u_0 \tau_R$.}
\label{fig:1}
\end{figure*}

Consider a dilute suspension of noninteracting, active spherical particles of mass $m$ and
radius $a$ propelling in a quiescent, homogeneous, isotropic, viscoelastic solvent in a stationary state, $\psi_0 = \left(\rho_0, \mathbf{v_0} = \mathbf{0}, s_0 \right)$, characterized by the local mass density, $\left(\rho_0 \right)$, hydrodynamic velocity, $\left(\mathbf{v_0} \right)$, and entropy density, $\left(s_0 \right)$. The solvent is a layer of a linear viscoelastic fluid with thickness $d$ and confined between two parallel (solid) plates perpendicular to the $z$-axis, but unbounded in the $x$ and $y$ directions, as depicted in Fig.~\ref{fig:1}.

\subsection{The solvent}

The initial quiescent state $\psi_0$ is altered by the self-propelled motion of the particles. Since active particles are intrinsically out of equilibrium, this results in a NESS. If the characteristic propulsion velocity of the active particles is not too high, the relaxation dynamics of the surrounding viscoelastic fluid can be described using linear response (LR) theory. Furthermore, if the temperature perturbations in these processes are sufficiently small, its density fluctuations originate only in pressure fluctuations. If $\mathbf{v_0}$ is much smaller than the sound velocity in the fluid, pressure fluctuations can be neglected, and the solvent may be considered as incompressible \cite{landau1959}. For arbitrary states, the mass, $\rho(\mathbf{r},t)$, and entropy, $s(\mathbf{r},t)$, densities (per unit volume), of the solvent form a set of three hydrodynamic variables, $\psi(\mathbf{r},t)=(\rho,\mathbf{v},s)$, which do not couple to any other variables.

In the LR regime, the most general constitutive equation for the linear stress-strain relation for a linear, homogeneous, viscoelastic fluid, is of the general form \cite{ferry1980,wang1980}
\begin{widetext}
\begin{equation}\label{eq:stress}
    \sigma_{ij}(\mathbf{r},t) = -p\delta_{ij} + \int_0^t dt' \left\{K(t-t') \dot{\gamma}_{kk}(\mathbf{r},t')\delta_{ij} + 2G(t-t')\left[ \dot{\gamma}_{ij}(\mathbf{r},t') - \frac{1}{3}\dot{\gamma}_{kk}({\mathbf{r}},t') \delta_{ij} \right] \right\},
\end{equation}
\end{widetext}
where $p(\mathbf{r},t)$ denotes the local pressure and $2\dot{\gamma}_ {ij}(\mathbf{r} ,t) \equiv \frac{\partial v_i}{\partial x_j} + \frac{\partial v_j}{\partial x_i}$ is the strain-rate tensor. In this linear homogeneous viscoelastic approximation, the bulk
compressional modulus $K(t)$ and the shear modulus $G(t)$ are scalar and spatially independent functions. Hence, under isothermal conditions, the solvent is incompressible, \emph{i.e.}, $\rho(\mathbf{r},t) = \rho_0$. In addition, if the characteristic propulsion speed of the particles is sufficiently slow and the suspension is so diluted that the presence of the active particles does not appreciably perturb the motion of the fluid, the dynamics of the solvent is described by the continuity equation
\begin{equation}\label{eq:continuity}
    \nabla \cdot \mathbf{v}(\mathbf{r},t) = 0,
\end{equation}
and the equation of motion
\begin{widetext}
\begin{equation}\label{eq:motion}
    \rho_0 \left[ \frac{\partial}{\partial t} + \mathbf{v}(\mathbf{r},t) \cdot \nabla \right] \mathbf{v}(\mathbf{r},t) = -\nabla p(\mathbf{r},t) + \int_0^t dt' \left\{ \left[ K(t-t') + \frac{1}{3}G(t-t') \right] \nabla \nabla \cdot \mathbf{v} (\mathbf{r},t') + G(t-t') \nabla^2 \mathbf{v}(\mathbf{r},t') \right\}. 
\end{equation}
\end{widetext}

\subsection{The active particles}

If no chemical reactions occur between the particles, their total mass is conserved and their local number density, $n(\mathbf{r},t)$, obeys a generalized Fick’s law \cite{alley1979,bonet1990}
\begin{equation}\label{eq:genFick}
    \mathbf{J}(\mathbf{r},t) = - \int_0^t dt' D(t-t') \nabla n(\mathbf{r},t') + n(\mathbf{r},t) \mathbf{v}(\mathbf{r},t),
\end{equation}
where the flux of active particles, $\mathbf{J}(\mathbf{r},t)$, displays time-memory effects in the diffusion
coefficient $D(t-t')$ of the particles due to the viscoelasticity of the solvent. $D(t-t')$ depends on
times $t'$ previous to the observation time $t$. The associated generalized diffusion equation governing the time evolution of $n(\mathbf{r},t)$ is
\begin{equation}\label{eq:gendiffusion}
    \frac{\partial}{\partial t} n(\mathbf{r},t) + \mathbf{v}(\mathbf{r},t) \cdot \nabla n(\mathbf{r},t) = \int_0^t D(t-t') \nabla^2 n(\mathbf{r},t'), 
\end{equation}
where the second term on the l.h.s. denotes the convective flux of particles arising from the
flow of the solvent. In section \ref{sect:diff} we propose a model to obtain an explicit analytic expression
for $D(t)$ in Eq.~(\ref{eq:genFick}).

\subsection{Stationary states}\label{subsect:NESS}

Once $G(t)$ and $D(t)$ are known, Eqs. (\ref{eq:stress})-(\ref{eq:gendiffusion}) form a complete system of hydrodynamic equations for the active suspension. The stationary (time independent) solutions of this system define nonequilibrium steady states (NESS) determined by the boundary conditions imposed on the system. Here we shall only consider the stationary solution corresponding to a uniform number density of active particles
\begin{equation}\label{eq:grad}
    n(\mathbf{r},t)  = n_0,
\end{equation}
\emph{i.e.}, $\nabla n(\mathbf{r},t) = \mathbf{0}$, with $\mathbf{v}_0(\mathbf{r}) = \mathbf{0}$ and constant mass density and pressure of the solvent. Extensions of the NESS situations studied here might include spatial variations in the number density, \emph{i.e}, $n(\mathbf{r},t) = n_{st}(\mathbf{r})$ with $\nabla n_{st}(\mathbf{r}) \neq \mathbf{0}$.

\section{Fluctuations}\label{sect:fluct}

We now introduce thermal fluctuations into Eqs. (\ref{eq:continuity}), (\ref{eq:motion}) and (\ref{eq:gendiffusion}) based on Landau’s fluctuating
hydrodynamics \cite{landau1959,ortiz2006}. To be consistent with LR, we consider small fluctuations $\delta \psi_s(\mathbf{r},t) = \psi(\mathbf{r},t) - \psi_0$, around the steady-state, with $\delta \mathbf{v}(\mathbf{r},t)= \mathbf{v}(\mathbf{r},t)$, due to Galilean invariance. Formally, this is accomplished by adding a momentum fluctuating source $\mathbf{\Pi}(\mathbf{r},t)$ into (\ref{eq:motion}), and a stochastic current $\mathbf{J}(\mathbf{r},t)$ to the flux into (\ref{eq:gendiffusion}). By linearizing the resulting equations in $\delta \psi_s$, we arrive at the following complete set of linearized equations for the fluctuations
\begin{widetext}
\begin{equation}\label{eq:fluctincompress}
    \nabla \cdot \delta \mathbf{v}(\mathbf{r},t) = 0
\end{equation}
\begin{equation}\label{eq:fluctmotion}
    \rho_0 \frac{\partial}{\partial t} \delta \mathbf{v} (\mathbf{r},t)  = -\nabla \delta p + \int_0^t dt' G(t-t') \nabla^2 \delta \mathbf{v}(\mathbf{r},t') + \nabla \cdot \mathbf{\Pi}(\mathbf{r},t) 
\end{equation}
\begin{equation}\label{eq:fluctgendiffusion}
    \frac{\partial}{\partial t} \delta n(\mathbf{r},t) =  \int_0^t dt' D(t-t') \nabla^2 \delta n(\mathbf{r},t') - \nabla \cdot \mathbf{J}(\mathbf{r},t).
\end{equation}
\end{widetext}
Eq. (\ref{eq:fluctmotion}) can be further simplified by applying the operator $\nabla \times \nabla \times$
and using the incompressibility condition (\ref{eq:fluctincompress}), thus eliminating the gradient term with the result
\begin{widetext}
\begin{equation}\label{eq:fluctmotion2}
    \rho_0 \frac{\partial}{\partial t} \nabla^2 \delta \mathbf{v}(\mathbf{r},t) = \int_0^t dt' G(t-t') \nabla^2 \nabla^2 \delta \mathbf{v}(\mathbf{r},t') + \nabla^2\left[\nabla \cdot \mathbf{\Pi}(\mathbf{r},t) \right] - \nabla \left\{ \nabla \cdot \left[ \nabla \cdot \mathbf{\Pi}(\mathbf{r},t)\right] \right\}.
\end{equation}
\end{widetext}
The random terms $\mathbf{J} (\mathbf{r} ,t)$ and $\mathbf{\Pi} (\mathbf{r} ,t)$ in (\ref{eq:fluctmotion}) and (\ref{eq:fluctgendiffusion}) are modeled as Gaussian, stationary, non-Markovian stochastic processes with zero mean \cite{dorfman1994}
\begin{equation}\label{eq:meannoise}
    \langle \mathbf{\Pi} (\mathbf{r} ,t) \rangle = \mathbf{0}, \,\,\,\,\, \langle \mathbf{J} (\mathbf{r} ,t) \rangle = \mathbf{0},
\end{equation}
and correlations given by the following relations derived from the expression for the rate of change of the total entropy of the fluid  \cite{ortiz2006,dorfman1994}
\begin{equation}\label{eq:correlmomentum}
    \langle \Pi_{ij}(\mathbf{r} ,t) \Pi_{lm}(\mathbf{r'} ,t') \rangle = 2k_B T G(|t-t'|) \delta(\mathbf{r}-\mathbf{r'})\Delta_{ijlm}
\end{equation}
\begin{equation}\label{eq:correlcurrent}
    \langle J_i(\mathbf{r} ,t) J_j(\mathbf{r'} ,t') \rangle = 2n_0 D(|t-t'|) \delta(\mathbf{r}-\mathbf{r'})\delta_{ij},
\end{equation}
where the angular brackets denote an average over the NESS, $k_B$ is the Boltzmann constant
and the tensor $\Delta_{ijlm}$ is defined by
\begin{equation}\label{eq:Delta}
    \Delta_{ijlm} = \delta_{ij} \delta_{lm} + \delta_{im} \delta_{jl} - \frac{2}{3} \delta_{ij} \delta_{lm},
\end{equation}
where $\delta_{ij}$ denotes the Kronecker delta.

Since a description in terms of the diffusion coefficient is expected to be valid only at times much longer than the molecular times, the upper limit in Eqs. (\ref{eq:fluctgendiffusion}) and (\ref{eq:fluctmotion2}) may be extended to $+\infty$, and their Fourier transform can be used,
\begin{equation}\label{eq:Fouriertransform}
    \hat{A}(\mathbf{k},\omega) = \int d\mathbf{r} \int_{-\infty}^{+\infty} dt \, e^{i \mathbf{k} \cdot \mathbf{r}} e^{-i\omega t} A(\mathbf{r},t).
\end{equation}
Then, the fluctuating linearized Eqs. (\ref{eq:fluctincompress}), (\ref{eq:fluctgendiffusion}), (\ref{eq:fluctmotion2})
in $(\mathbf{k},\omega)$ space read
\begin{equation}\label{eq:Fourierincompress}
    \mathbf{k} \cdot \delta \hat{\mathbf{v}}(\mathbf{k},\omega) = 0,
\end{equation}
\begin{equation}\label{eq:Fouriergendiffusion}
    \delta \hat{n}(\mathbf{k}.\omega) = \hat{g}(\mathbf{k}.\omega)  i \mathbf{k} \cdot \hat{\mathbf{J}}(\mathbf{k},\omega).
\end{equation}
\begin{equation}\label{eq:Fouriermotion}
    \delta \hat{\mathbf{v}}(\mathbf{k},\omega) = -\hat{h}(\mathbf{k},\omega) \left( \mathbb{I} - \mathbf{k} \mathbf{k}\right) \cdot \left[ i\mathbf{k}\cdot \hat{\mathbf{\Pi}}(\mathbf{k},\omega)  \right],
\end{equation}
where the Green functions associated with Eqs. (\ref{eq:fluctgendiffusion}) and (\ref{eq:fluctmotion2}) are, respectively,
\begin{equation}\label{eq:Greeng}
    \hat{g}(\mathbf{k},\omega) \equiv \frac{1}{i\omega + k^2 \hat{D}(\omega)}
\end{equation}
\begin{equation}\label{eq:Greenh}
    \hat{h}(\mathbf{k},\omega) \equiv \frac{1}{i \rho_0 \omega + k^2 \hat{G}(\omega)},
\end{equation}
being $k = |\mathbf{k}|$. In Fourier space the relations (\ref{eq:correlmomentum}) and (\ref{eq:correlcurrent}) then become
\begin{widetext}
\begin{equation}\label{eq:Fouriercorrelmomentum}
   \left\langle \hat{\Pi}_{ij}(\mathbf{k},\omega) \hat{\Pi}_{lm}(\mathbf{k'},\omega')  \right\rangle = 4 (2\pi)^4 k_B T  \mathfrak{Re} \left[ \hat{G}(\omega)\right] \delta(\mathbf{k}+\mathbf{k'}) \delta(\omega + \omega') \Delta_{ijlm},
\end{equation}
\begin{equation}\label{eq:Fouriercorrelcurrent}
   \left\langle \hat{\mathbf{J}}(\mathbf{k},\omega) \hat{\mathbf{J}}(\mathbf{k}',\omega') \right\rangle = 4 (2\pi)^4 n_0 \mathfrak{Re} \left[ \hat{D}(\omega)\right] \delta(\mathbf{k}+\mathbf{k}') \delta(\omega+\omega')  \mathbb{I},
\end{equation}
\end{widetext}
where $\mathbb{I}$ is the unit tensor and $\delta(\mathbf{k}+\mathbf{k}')$ and $\delta(\omega+\omega')$ stand for the Dirac delta function in wave vector and frequency domains, respectively.

\section{Active diffusion model}\label{sect:diff}

It should be stressed that in the above equations, the explicit form of the generalized diffusion coefficient $D(t)$ is valid for any form of $\hat{D}(\omega)$, however, its explicit form should be consistent with the active character of the self-propelled particles and with the specific type of the solvent viscoelasticity. In this section, we propose a model to obtain an explicit expression for $D(t)$. As sketched in Fig. \ref{fig:1}(b), we model the motion of a spherical active particle by a generalized Langevin equation of the form \cite{narinder2018,saad2019,lozano2019,muhsin2021,sprenger2022}
 \begin{equation}\label{eq:GLEactive}
    m\frac{d}{dt}\mathbf{u}(t) = - \int_0^t \gamma(t-t') \left[ \mathbf{u}(t') - \mathbf{u}_{sp}(t') \right]dt' + \bm{\zeta}_T(t),
 \end{equation}
Here, $\mathbf{u}(t) = \frac{d}{dt} \mathbf{r}(t)$ is the instantaneous velocity of the active particle, whose position at time $t$ is $\mathbf{r}(t)$, and the left hand side represents the inertial force. The first term on the right hand side is the net friction acting on a particle that is self-propelling at velocity $\mathbf{u}_{sp}(t)$, and $\bm{\zeta}_T(t)$ is assumed to be a Gaussian noise which mimics the effects of thermal fluctuations,
\begin{equation}\label{eq:thermalnoise}
    \langle \bm{\zeta}_T(t) \rangle = \bm{0}, \,\,\,\,\, \langle \bm{\zeta}_T(t) \bm{\zeta}_T(t') \rangle = k_B T \gamma(|t-t'|) \mathbb{I} .
\end{equation}
 For the sake of simplicity, we focus on the overdamped limit 
\begin{equation}\label{eq:overdampedGLEactive}
    \int_0^t \gamma(t-t') \mathbf{u}(t') dt' = \int_0^t \gamma(t-t') \mathbf{u}_{sp}(t') dt' + \bm{\zeta}_T(t),
\end{equation}
and assume that the propulsion velocity $\mathbf{u}_{sp}(t)$ corresponds to an Ornstein-Uhlenbeck process
\begin{equation}\label{eq:OU}
    \frac{d}{dt} \mathbf{u}_{sp}(t) = - \frac{1}{\tau_R} \mathbf{u}_{sp}(t) + \bm{\xi}_{sp}(t),
\end{equation}
where $\bm{\xi}_{sp}(t)$ is a Gaussian white noise satisfying 
\begin{equation}\label{eq:activewhitenoise}
\langle \bm{\xi}_{sp}(t) \rangle = \mathbf{0}, \,\,\,\,\, \langle \bm{\xi}_{sp}(t) \bm{\xi}_{sp}(t') \rangle = 2\frac{u_0^2}{\tau_R}\delta(t-t') \mathbb{I}, 
\end{equation}
with $u_0$ representing the characteristic propulsion speed. Moreover, $\bm{\xi}_{sp}$ is decorrelated from $\zeta_T(t)$, \emph{i.e.}, 
\begin{equation}\label{eq:decorractivethermal}
\langle \bm{\xi}_{sp}(t) \bm{\zeta}_T(t') \rangle = \mathbb{O}, 
\end{equation}
where $\mathbb{O}$ is the zero tensor. Eq. (\ref{eq:OU}) captures qualitatively the effect of rotational diffusion at sufficiently long times  with effective rotational diffusion time $\tau_R$ or tumbling rate $\tau_R^{-1}$ in the persistence of self-propulsion with characteristic propulsion speed $u_0$, as observed in numerous experiments and simulations of self-propelled particles in complex media \cite{narinder2018,lozano2019,saad2019,gomezsolano2016,aragones2018,yuan2019,theeyancheri2020,qi2020}. The general solution of equation (\ref{eq:overdampedGLEactive}) is
\begin{widetext}
\begin{equation}\label{eq:solvel}
    \mathbf{u}(t) = \mathbf{u}_0\exp\left(-\frac{t}{\tau_R}\right) + \int_0^t \exp\left( -\frac{t-t'}{\tau_R} \right) \bm{\xi}_{sp}(t')dt' + \int_0^t \Gamma(t-t') \bm{\zeta}_T(t')dt',
\end{equation}
\end{widetext}
where $\Gamma(t-t')$ corresponds to the inverse Laplace transform of $\frac{1}{\tilde{\gamma}(\epsilon)}$, being $\tilde\gamma(\epsilon) = \int_0^{\infty} e^{-\epsilon t} \gamma(t) dt$ the Laplace transform of the friction kernel $\gamma(t)$. Therefore, for $t \ge t'  \gg \tau_R$, the velocity autocorrelation function of the active particle in a NESS can be written as 
\begin{widetext}
\begin{eqnarray}\label{eq:autocorrvel}
    \langle \mathbf{u}(t) \mathbf{u}(t') \rangle & =  &
        \int_0^t ds\int_0^{t'} ds' e^{-\frac{t-s}{\tau_R}} e^{-\frac{t'-s'}{\tau_R}} \langle \bm{\xi}_{sp}(s) \bm{\xi}_{sp}(s') \rangle 
         +  \int_0^t ds\int_0^{t'} ds' \Gamma(t-s) \Gamma(t'-s') \langle \bm{\zeta}_T(s) \bm{\zeta}_{T}(s') \rangle ,\nonumber\\
        & = & u_0^2 e^{ -\frac{t - t'}{\tau_R} } \mathbb{I} +  \int_0^t ds\int_0^{t'} ds' \Gamma(t-s) \Gamma(t'-s') \langle \bm{\zeta}_T(s) \bm{\zeta}_{T}(s') \rangle.
\end{eqnarray}
\end{widetext}
On the other hand, according to the definition of the time-dependent diffusion function of the generalized Fick's law, see equation (54) in \cite{rodriguez2013}
\begin{equation}\label{eq:gendiff}
    D(t) = \frac{1}{3}\mathrm{Tr}\left[\langle \mathbf{u}(t) \mathbf{u}(0) \right] \rangle \Theta(t),
\end{equation}
where $\Theta(t)$ is the Heaviside step function. Thus, from equation (\ref{eq:autocorrvel}) it can be seen that $D(t)$ can be written as the sum of two contributions, $D(t) = D_T(t) + D_{sp}(t)$: a thermal contribution
$D_T(t)$, whose Fourier transform can be approximated in the overdamped limit as
\begin{equation}\label{eq:Fourierthermaldiff}
    \hat{D}_T(\omega ) = \int_{-\infty}^{\infty} dt \, e^{-i\omega t} D_{T}(t) = \frac{k_B T}{\hat{\gamma}(\omega)},
\end{equation}
where $\hat{\gamma}(\omega)$ is the Fourier transform of the memory kernel in Eq. (\ref{eq:overdampedGLEactive})
as computed in \cite{rodriguez2013}, whereas the active contribution 
\begin{equation}\label{eq:actdiff}
    D_{sp}(t) = u_0^2 e^{-\frac{t}{\tau_R}}\Theta(t),
\end{equation}
can be expressed in frequency domain as 
\begin{equation}\label{eq:Fourieractdiff}
    \hat{D}_{sp}(\omega)  = \frac{u_0^2}{\frac{1}{\tau_R} + i\omega}.
\end{equation}
Therefore, in the limit of very large penetration length \cite{indei2012} the simplest frequency-dependent diffusion function involved in the generalized Fick's equation which accounts for both the effect of the viscoelastic solvent and the persistent motion of the active particle can me modeled as
\begin{equation}\label{eq:frequencydiff}
    \hat{D}(\omega) = \frac{k_B T}{6 \pi a \hat{G}(\omega)} + \frac{u_0^2}{\frac{1}{\tau_R} + i\omega},
\end{equation}
where $\hat{G}(\omega)$ is the Fourier transform of the stress relaxation modulus of the viscoelastic solvent. 
Note that a description in terms of the frequency-dependent diffusion coefficient given by Eq. (\ref{eq:frequencydiff}) is expected to be valid for times much longer than molecular times. Indeed, for long timescales of the order of the relaxation times of the diffusion modes, the leading contributions to the thermal diffusive term, $\hat{D}_T(\omega)$ given by Eq. (\ref{eq:Fourierthermaldiff}), are determined by the viscoelastic nature of the solvent, which is fully characterized by $\hat{G}(\omega)$. This is in turn related to the dynamic shear modulus $G^{*}(\omega)$, a quantity commonly measured in small-amplitude oscillatory shear rheology, by means of
\begin{eqnarray}\label{eq:shearmodulus}
    G^{*}(\omega) & \equiv & G'(\omega) + i G''(\omega)\nonumber\\
    & = & i\omega \hat{G}(\omega),
\end{eqnarray}
where $G'(\omega)$ and $G''(\omega)$ are the storage and loss modulus, respectively \cite{ferry1980}. 
On the other hand. the effective description of $\hat{D}_{sp}(\omega)$ given by Eq. (\ref{eq:Fourieractdiff}) is valid only for timescales that are much larger than those of the molecular mechanisms responsible for the randomization of the particle's persistent direction. Therefore, under such conditions, Eq. (\ref{eq:frequencydiff}) must represent an adequate model for the effective diffusive behavior of run-and-tumble particles \cite{schnitzer1993}, active Brownian particles \cite{hagen2011}, as well as Ornstein-Uhlenbeck particles \cite{fily2012}, moving in a viscoelastic medium.

\section{Dynamic structure factor}\label{sect:dsf}

In this section, we calculate a general expression for the dynamic structure factor of the active suspension, which is simply given by the spectrum of the number-density fluctuations of active particles in the solvent in the NESS described in Subsect. \ref{subsect:NESS}. Thus, from Eqs. (\ref{eq:Fouriergendiffusion}), (\ref{eq:Greeng}), and (\ref{eq:Fouriercorrelcurrent}), the dynamic structure factor is given by
\begin{eqnarray}\label{eq:SF0}
   \hat{S}\left(\mathbf{k},\omega \right) & \equiv & \left\langle \delta \hat{n}(\mathbf{k},\omega) \delta \hat{n}(-\mathbf{k},-\omega)\right\rangle \nonumber\\
   & = & \frac{4 (2\pi)^4 n_0 \delta^4(0) k^2\mathfrak{Re} \left[ \hat{D}(\omega) \right]}{\left\{ \omega + k^2 \mathfrak{Im} \left[ \hat{D}(\omega) \right] \right\}^2 + k^4 \left[ \mathfrak{Re} \left[ \hat{D}(\omega) \right] \right]^2}.
\end{eqnarray}
We now consider specific models for the dyanmic shear modulus of the medium, $G^*(\omega)$, and analyze the main effects on the resulting dynamic structure factor for distinct values of the parameters that characterize the dilute active-particle suspension. 

\subsection{Newtonian solvent}\label{subsect:Newtonian}

\begin{figure*}\centering
 \includegraphics[width=0.875\textwidth]{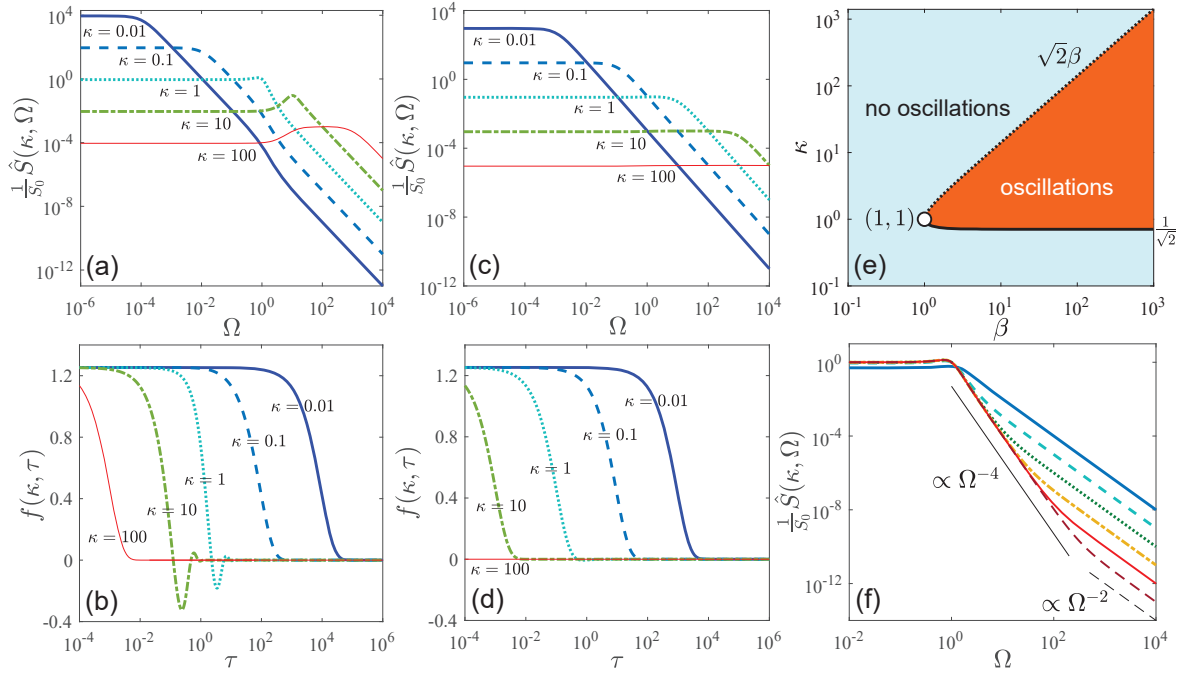}
 \caption{(a) Dynamic structure factor as a function of the normalized frequency $\Omega = \tau_R \omega$, and (b) corresponding intermediate scattering function as a function of the normalized time $\tau=t/\tau_R$ of an active suspension in a Newtonian solvent with $\beta = 10$ and distinct values of the normalized wave number $\kappa = u_0 \tau_R k$. (c) Dynamic structure factor as a function of the normalized frequency $\Omega$, and (d) corresponding intermediate scattering function as a function of the normalized time $\tau$ of an active suspension in a Newtonian solvent with $\beta = 0.1$ and distinct values of $\kappa$. (e) Diagram of the different regimes of the dynamical structure of the active suspension with Newtonian solvent that can be probed depending on the values of $\beta$ and $\kappa$. The values $\beta = 1$ and $\kappa=1$ (white circle) mark the onset of oscillations in the intermediate scattering function, which are intrinsic to the active nature of the particles. The solid and dashed lines represent the characteristic wave numbers $\kappa_-$ and $\kappa_+$ given by Eq. (\ref{eq:kappa}), respectively. (f) Dynamic structure factor as a function of the normalized frequency $\Omega$ for $\kappa = 1$ and different values of $\beta$: $ 1$ (thick solid line), 10 (thick dashed line), $10^2$ (thick dotted line), $10^3$ (thick dotted-dashed line), $10^4$ (thin solid line), and $10^5$ (thin dashed line). The thinner black solid and dashed lines depict the behaviors $\propto \Omega^{-4}$ and $\propto \Omega^{-2}$, respectively.}
\label{fig:2}
\end{figure*}

In this case, the storage and loss modulus in Eq. (\ref{eq:shearmodulus}) are given by
\begin{equation}\label{eq:Newtonianstorage}
    G'(\omega) = 0,
\end{equation}
\begin{equation}\label{eq:Newtonianloss}
    G''(\omega) = \eta \omega,
\end{equation}
respectively, where $\eta$ is the constant (frequency-independent) shear viscosity of the Newtonian liquid. In such a case, the dynamic structure factor (\ref{eq:SF0}) is explicitly given by the expression
\begin{equation}\label{eq:dynstructfactNewton}
     \hat{S}(\bm{\kappa},\Omega) = S_0 \frac{\beta \kappa^2 \left(1 + \frac{\beta}{1 + \Omega^2}  \right)}{\beta^2 \left( 1 - \frac{\kappa^2}{1+\Omega^2}  \right)^2 \Omega^2 + \kappa^4 \left( 1 + \frac{\beta}{1 + \Omega^2}  \right)^2},
\end{equation}
where $S_0 = 4(2\pi)^4n_0 \delta^4(0) \tau_R$, and we have chosen the rotational time, $\tau_R$, and the persistence length, $u_0 \tau_R$, of the active particles to define the dimensionless frequency, $\Omega = \tau_R \omega$, and the dimensionless wave vector, $\bm{\kappa} = u_0 \tau_R \mathbf{k}$, respectively, with  $\kappa = |\bm{\kappa}|$. In addition, in Eq. (\ref{eq:dynstructfactNewton}) we have introduced the dimensionless parameter
$\beta = \frac{u_0^2 \tau_R}{\frac{k_B T}{6\pi a \eta}} = \frac{D_A}{D_T}$, where $D_A = u_0^2 \tau_R$ is the active diffusion coefficient of the particles, and $D_T = \frac{k_B T}{6\pi a \eta}$ is their long-time thermal diffusion coefficient, in order to quantify the effect of self-propulsion relative to thermal fluctuations. Note that, for $u_0 = 0$, Eq. (\ref{eq:dynstructfactNewton}) reduces to the Lorentzian function $\hat{S}(\mathbf{k},\omega)\propto \frac{\omega_c}{\omega^2 + \omega_c^2}$ with cut-off frequency $\omega_c = D_T k^2$ and zero-frequency plateau $\hat{S}(\mathbf{k},\omega \rightarrow 0) \propto \omega_c^{-1}$ that describes the equilibrium dynamics of a dilute suspension of passive Brownian particles in a viscous solvent for all $\mathbf{k}$. In this case, the inverse Fourier transform  corresponds to the well-known exponential behavior of the intermediate scattering function, $F(\mathbf{k},t)  \propto \exp \left( - D_T k^2 t \right) $. Nevertheless, self-propulsion of the particles in the solvent strongly modifies the shape of the dynamic structure factor of the suspension, as can be seen in Fig. \ref{fig:2}(a) for $\beta = 10$. For instance, for $\kappa = 0.01$ and 0.1, \emph{i.e.}, for wave numbers smaller than the inverse of the persistence length, a Lorentzian-like behavior is observed for sufficiently small frequencies ($\Omega < 1$). Indeed, it can be easily checked from Eq. (\ref{eq:dynstructfactNewton}) that, for $\kappa \ll 1$, the dynamic structure factor as a function of $\omega$ can be approximated to a Lorentzian curve with cut-off frequency $\omega_c = \left( D_T + D_A \right)k^2$, which corresponds to wave numbers at which the active suspension effectively behaves as a passive suspension of Brownian particles with an effective diffusion coefficient $D_T + D_A$. This behavior stems from the fact that, at timescales that are much longer than $\tau_R$, active particles lose the persistence in their self-propelled motion, thus effectively performing a random walk. However, in contrast to purely passive suspensions, as the wave number $k$ becomes comparable or larger than $\left( u_0 \tau_R \right)^{-1}$, systematic deviations from the Lorentzian profile occur. A particular feature is the emergence of a well-defined peak around a given frequency, whose location along the $\Omega$-axis and characteristic width both increase as $\kappa$ increases, as shown in Fig. \ref{fig:2}(a) for $\kappa = 1,10$ and 100. Such a peak is the hallmark of the particle activity in the dynamical structure of the suspension, which results from the persistent motion probed at length-scales that are similar or smaller than the persistence length $u_0 \tau_R$. To better appreciate in time domain the role of such a peak, we compute the normalized intermediate scattering function, defined as
\begin{equation}\label{eq:ISF}
    f(\bm{\kappa},\tau) =  \frac{1}{2\pi S_0} \int_{-\infty}^{\infty} d\Omega \, e^{i\Omega \tau} \hat{S}(\mathbf{\kappa},\Omega).
\end{equation}
In Fig. \ref{fig:2}(b) we plot the dependence of $f(\bm{\kappa},\tau)$ on the dimensionless time $\tau = t/\tau_R$, corresponding to the different structure-factor curves shown in Fig. \ref{fig:2}(a). From  the poles of Eq. (\ref{eq:dynstructfactNewton}), it can be readily shown that, for $\beta \ge 1$ (large particle activity) and $0 < \kappa < \kappa_-$ [see Eq. (\ref{eq:kappa})], $f(\bm{\kappa},\tau)$ is given by the sum of purely decaying exponentials, as can be verified by the monotonically decreasing dependence of $f(\bm{\kappa},\tau)$ on $\tau$ for $\kappa = 0.01$ and 0.1 shown in Fig. \ref{fig:2}(b). Moreover, for $\beta \ge 1$ and $\kappa_- \le \kappa \le \kappa_+$, where the characteristic wave number $\kappa_{\pm}$ are given by
\begin{equation}\label{eq:kappa}
    \kappa_{\pm} = \sqrt{\beta}\sqrt{\beta \pm \sqrt{\beta^2-1}},
\end{equation}
an oscillatory behavior of the intermediate scattering function emerges as a function of $\tau$, which is similar to the oscillations found at intermediate wave numbers for other models of active particles in a viscous environment \cite{kurzthaler2018,kurzthaler2016,kurzthaler2017,croze2019}. This can be checked in Fig. \ref{fig:2}(b), where oscillations in $f(\bm{\kappa},\tau)$ develop for the wave numbers $\kappa = 1$ and 10, which are contained in the interval $ \kappa_- = 0.708 < \kappa < \kappa_+ = 14.1244$. This reveals the peculiarities of particle self-propulsion in the non-equilibrium dynamics of the active suspension in a Newtonian solvent, which can be probed at intermediate wave numbers, and are absent in the case of passive particles. Furthermore, for $\beta \ge 1$ and $\kappa > \kappa_+$, even though a peak in the structure factor persists, the oscillations of the intermediate scattering function vanish, thus exhibiting a sum of exponentially decaying relaxations, as can be seen in Fig. \ref{fig:2}(b) for $\kappa = 100 > \kappa_+ = 14.1244$. On the other hand, for $0 < \beta < 1$ (small particle activity), qualitatively different behaviors of $\hat{S}(\mathbf{\kappa},\Omega)$ and  $f(\bm{\kappa},\tau)$, are found, as plotted in Figs. \ref{fig:2}(c) and \ref{fig:2}(d), respectively. In this regime, the dynamic structure factor exhibits a quasi-Lorentzian profile as a function of $\Omega$ for all values of $\kappa$, as verified in Fig. \ref{fig:2}(c) for $\kappa = 0.01,0.1,1,10$ and 100. This results in intermediate scattering functions that are given by the sum of exponential decays without any oscillatory behavior, as shown in Fig. \ref{fig:2}(d) for the same values of $\kappa$. 
Fig. \ref{fig:2}(e) summarizes the different regimes that can be probed in the dynamical structure of the active suspension depending on the specific values of $\beta$ and $\kappa$, where the values $\beta = 1$ and $\kappa = 1$ correspond to the onset of oscillatory behavior in the intermediate scattering function. In particular, it should be noted that, in the limit when active diffusion fully dominates over thermal diffusion, $\beta \gg 1$, oscillations in $f(\bm{\kappa},\tau)$ occur for wave numbers in the interval $\frac{1}{\sqrt{2}}  \le \kappa \lesssim  \sqrt{2}\beta$.

To illustrate another distinctive feature of the dynamical  structure of the suspension in the regime where oscillations in the intermediate scattering function emerge, in Fig. \ref{fig:2}(f) we plot $\hat{S}(\mathbf{\kappa},\Omega)$ as a function of $\Omega$ for $\kappa = 1$, \emph{i.e.}, at a wave number equal to the inverse of the persistence length $u_0 \tau_R$, and different values of $\beta \ge 1$. For $\Omega \gg \sqrt{\beta}$, we observe the behavior $\hat{S}(\bm{\kappa},\Omega) \approx S_0 \beta^{-1} \kappa^2 \Omega^{-2} $, which is expected  regardless of the values of $\kappa$ and $\beta$ due to the presence of thermal noise in the fluid, In addition, we uncover the emergence of an intermediate frequency range, $1 \lesssim \Omega \lesssim \sqrt{\beta}$, where $\hat{S}(\bm{\kappa},\Omega) \approx S_0 \kappa^2\Omega^{-4}$, which becomes broader as $\beta$ increases, as observed in Fig.~\ref{fig:2}(f) for $\beta = 10^0,10^1,10^2,10^3,10^4$ and $10^5$. This can be attributed to the active diffusion of the particles in the suspension, which becomes dominant with respect to purely thermal diffusion as $\beta$ increases.

\subsection{Maxwell-type solvent}

\begin{figure*}\centering
 \includegraphics[width=0.775\textwidth]{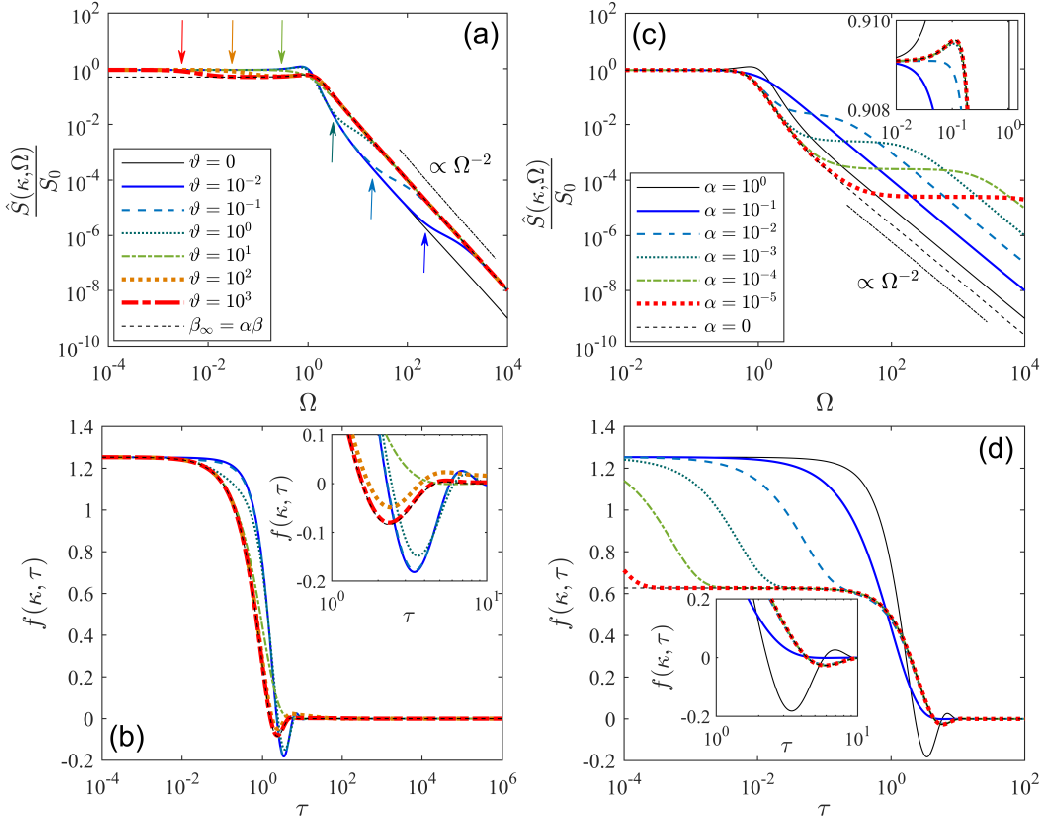}
 \caption{(a) Dynamic structure factor as a function of the normalized frequency $\Omega$, and (b) corresponding intermediate scattering function as a function of the normalized time $\tau$ of an active suspension with $\beta = 10$, probed at normalized wavenumber $\kappa = 1$, in a Maxwell-type viscoelastic solvent with $\alpha = 0.1$ and distinct values of the parameter $\vartheta$. The arrows in \ref{fig:3}(a) depict the frequencies $\Omega = \alpha^{-1/2} \vartheta^{-1}$ at which deviations from the Newtonian curve $(\vartheta = 0)$, start to develop. In \ref{fig:3}(b), the inset is an expanded view of the main plot around the values of $\tau$ at which the first oscillation in the intermediate scattering function shows up. Same colors and line styles as in \ref{fig:3}(a). (c) Dynamic structure factor as a function of the normalized frequency $\Omega$, and (d) corresponding intermediate scattering function as a function of the normalized time $\tau$ of an active suspension with $\beta = 10$ probed at normalized wave number $\kappa = 1$, in a Maxwell-type viscoelastic solvent with $\vartheta = 10$ and distinct values of the parameter $\alpha$. In \ref{fig:3}(c) and \ref{fig:3}(d), the insets are expanded views of the main plots around the values of $\Omega$ and $\tau$ at which the peak in $\hat{S}(\bm{\kappa},\Omega)$ and the first oscillation in $ f(\bm{\kappa},\tau)$ show up, respectively. Same colors and line styles as in \ref{fig:3}(c).}
\label{fig:3}
\end{figure*}

We now focus on a Maxwell-type solvent described by the following storage and loss modulus
\begin{equation}\label{eq:Maxwellstorage}
    G'(\omega) = \frac{\left( \eta_0 - \eta_{\infty} \right) \tau_0 \omega^2}{1 + \tau_0^2 \omega^2},
\end{equation}
\begin{equation}\label{eq:Maxwellloss}
    G''(\omega) = \eta_{\infty} \omega + \frac{\left( \eta_0 - \eta_{\infty} \right) \omega}{1 + \tau_0^2 \omega^2},
\end{equation}
respectively, where $\eta_0$, $\eta_{\infty}$ and $\tau_0$ represent the zero-shear viscosity, the high-frequency shear viscosity in the limit $\omega \gg \tau_0^{-1}$, and the single relaxation time of the fluid where the active particles self-propel. This rheological model describes adequately the linear viscoelastic response of the swimming medium used in several experiments with active suspensions of Janus colloids in aqueous polymer solutions \cite{gomezsolano2016,narinder2018,narinder2019,saad2019}. In such a case, the explicit expression of the dynamic structure factor (\ref{eq:SF0}) is
\begin{widetext}
\begin{equation}\label{eq:dynstructfactMaxwell}
     \hat{S}(\bm{\kappa},\Omega) = S_0 \frac{\beta \kappa^2 \left( \frac{1+\alpha \vartheta^2  \Omega^2}{1+\alpha^2 \vartheta^2  \Omega^2} + \frac{\beta}{1+\Omega^2}\right)}{
    \left\{ \beta \Omega + \kappa^2 \left[ \frac{ (1-\alpha)\vartheta \Omega}{1 + \alpha^2 \vartheta^2 \Omega^2} - \frac{\beta \Omega}{1+\Omega^2} \right]\right\}^2
    +\kappa^4 \left( \frac{1+\alpha \vartheta^2  \Omega^2}{1+\alpha^2 \vartheta^2  \Omega^2} + \frac{\beta}{1+\Omega^2}\right)^2 }.
\end{equation}
\end{widetext}
In Eq. (\ref{eq:dynstructfactMaxwell}), in addition to the parameter $\beta$ that quantifies the activity of the particles in the suspension, we have introduced the dimensionless parameters $\alpha = \frac{\eta_{\infty}}{\eta_0} \ge 0$, $\vartheta = \frac{\tau_0}{\tau_R} \ge 0$, which characterize the viscoelasticity of the fluid.  
Note that, when $\alpha \rightarrow 1$ or $\vartheta \rightarrow 0$, the dynamic shear modulus described by Eqs. (\ref{eq:Maxwellstorage}) and (\ref{eq:Maxwellloss}) tends to the Newtonian case, Eqs. (\ref{eq:Newtonianstorage}) and (\ref{eq:Newtonianloss}), respectively, with frequency-independent viscosity $\eta = \eta_{\infty} = \eta_0$, as described in Subsect. \ref{subsect:Newtonian}. Accordingly, in such a limit Eq. (\ref{eq:dynstructfactMaxwell}) reduces to Eq. (\ref{eq:dynstructfactNewton}). On the other hand, a strong viscoelastic behavior of the fluid corresponds to $\alpha \ll 1$ or $\vartheta \gg 1$, which represent situations in which the flow resistance of the fluid microsctructure is much higher at low deformation rates and its relaxation time is very large with respect to other time-scales, respectively. To understand the influence of such viscoelastic parameters on the dynamics of the number density fluctuations of the active suspension, we first compute the dynamic structure factor as a function of $\Omega$ for $\beta = 10$ (dominant active diffusion with respect to thermal diffusion), $\kappa = 1$,  $\alpha = 0.1$ (large zero-shear viscosity with respect to high-frequency shear viscosity) and distinct values of the ratio of timescales $\vartheta = \tau_0 / \tau_R$ spanning several orders of magnitude: $\vartheta = 10^{-2}, 10^{-1}, 10^{0}, 10^1, 10^2$ and $10^3$, see Fig. \ref{fig:3}(a). Note that, for such values of $\beta$ and $\kappa$, the persistent nature of the active particles in a Newtonian solvent ($\vartheta = 0$) is manifested through a peak in $\hat{S}(\bm{\kappa},\Omega)$ around $\Omega \approx 1$, immediately followed by a monotonic decay $\hat{S}(\bm{\kappa},\Omega) \approx S_0 \beta^{-1} \kappa^2 \Omega^{-2}$ for $\Omega \gtrsim \sqrt{\beta} = \sqrt{10}$ due to thermal diffusion, whereas a low-frequency plateau 
$\hat{S}(\bm{\kappa},\Omega \rightarrow 0) = S_0 \kappa^{-2} \beta (1+\beta)^{-1}$ is conspicuous for $\Omega \ll 1$,  see the thin black solid line in Fig. \ref{fig:3}(a). However, a non-zero relaxation time of the fluid leads to systematic deviations of the dynamic structure factor from the Newtonian case, which are due to the presence of a high-frequency viscosity $\eta_{\infty} < \eta_0$ (in this specific example: $\eta_{\infty} = \alpha \eta_0 = 0.1\eta_0$). For a given value $\vartheta > 0 $, such deviations result in a transition around $\Omega \approx \alpha^{-1/2} \vartheta^{-1}$ to a qualitatively distinct regime, which is depicted by the arrows in Fig. \ref{fig:3}(a). Depending on the specific value of $\vartheta$, three different types of profiles of $\hat{S}(\bm{\kappa},\Omega)$ are observed. For instance, for $\vartheta \lesssim 1$ (short fluid relaxation time as compared to the particle rotational diffusion time), the peak in the dynamic structure factor remains rather unaltered around $\Omega \approx 1$, whereas a second diffusive behavior $\hat{S}(\bm{\kappa},\Omega) \approx S_0 \beta_{\infty}^{-1} \kappa^2 \Omega^{-2}$ sets in for $\Omega \gtrsim \alpha^{-1} \vartheta^{-1}$, where $\beta_{\infty}  < \beta$ represents a high-frequency activity parameter.
On the other hand, for $\vartheta \approx 10$, the peak vanishes, which results in a Lorentzian-like shape of the structure factor. This is verified in Fig. \ref{fig:3}(b), where oscillations in the intermediate scattering function are actually suppressed for such a specific value of $\vartheta$. Finally, for $\vartheta \gtrsim 10$, the peak re-emerges but at a frequency that is higher than that for $\vartheta \lesssim 1$. In all cases with non-zero $\vartheta$, a low-frequency plateau $\hat{S}(\bm{\kappa},\Omega \rightarrow 0) = S_0 \kappa^{-2} \beta (1+\beta)^{-1}$ that is consistent with a Newtonian fluid of viscosity $\eta_0$ can be observed, whereas the effective diffusive behavior characterized by $\beta_{\infty}$ is revealed at sufficiently high frequencies. Note that the numerical values of the dynamic structure factor in such a high-frequency regime are $\alpha^{-1} = 10$ times larger than those for the active suspension in a Newtonian solvent of viscosity $\eta_0$ with the same values of $\kappa$, $\Omega$ and $\beta$.
This can be explained by the existence of a high-frequency thermal diffusion coefficient associated to $\eta_{\infty}$, which is in this case $ \frac{k_B T}{6\pi a \eta_{\infty}} =  \frac{k_B T}{6\pi a \alpha \eta_{0}} = 10 D_T$, as verified for all curves with $\vartheta > 0$ at sufficiently large $\Omega$ in Fig. \ref{fig:3}(a), thereby shifting the aforementioned transition to smaller and smaller frequencies with increasing values of $\vartheta$. Furthermore, from Eq. (\ref{eq:dynstructfactMaxwell}), it can be demonstrated that in the limit $\vartheta \rightarrow \infty$, the dynamic structure factor converges to the curve corresponding to an active suspension in a Newtonian solvent, as described by Eq. (\ref{eq:dynstructfactNewton}), where the effective activity parameter is explicitly given by $\beta_{\infty} = \alpha \beta$. This limit corresponds to the situation where the rotational diffusion of the active particles is so fast that they do not have enough time to undergo the full relaxation of the fluid while self-propelling, thus effectively experiencing a Newtonian medium with viscosity $\eta_{\infty}$.  This is verified in Fig. \ref{fig:3}(a) for $\vartheta = 10^2$ and $10^3$, where apart from the plateau at low frequencies $\Omega \lesssim \alpha^{-1/2} \vartheta^{-1}$, the structure-factor curves have fully converged to the limiting case $\vartheta \rightarrow \infty$ for $\Omega \gtrsim \alpha^{-1} \vartheta^{-1}$. Such a behavior of the dynamic structure factor of the active suspension in a Maxwell-type viscoelastic solvent is also reflected in the dependence of the corresponding intermediate scattering function on the normalized time $\tau$, as illustrated in Fig. \ref{fig:3}(b). In the inset of this figure it becomes more evident that the specific value of $\vartheta$ determines the period of the oscillation in $f(\bm{\kappa},\tau)$ due to the particle activity. For small values of $\vartheta$, typically $\vartheta \lesssim 1$, the oscillatory dependence resembles that of a Newtonian fluid of viscosity $\eta_0$, whereas for $\vartheta \gg 1$, it approaches that of a Newtonian fluid of viscosity $\eta_{\infty}$. Such distinct oscillating behaviors of $f(\bm{\kappa},\tau)$ are separated by intermediate values of $\vartheta$, where a monotonic decay is observed in this example at $\vartheta = 10$. On the other hand, regardless of the value of $\vartheta$, all intermediate scattering function curves converge at short time-scales to $f(\bm{\kappa},\tau \rightarrow 0) \rightarrow \sqrt{\frac{\pi}{2}}$.

\begin{figure*}\centering
 \includegraphics[width=0.9\textwidth]{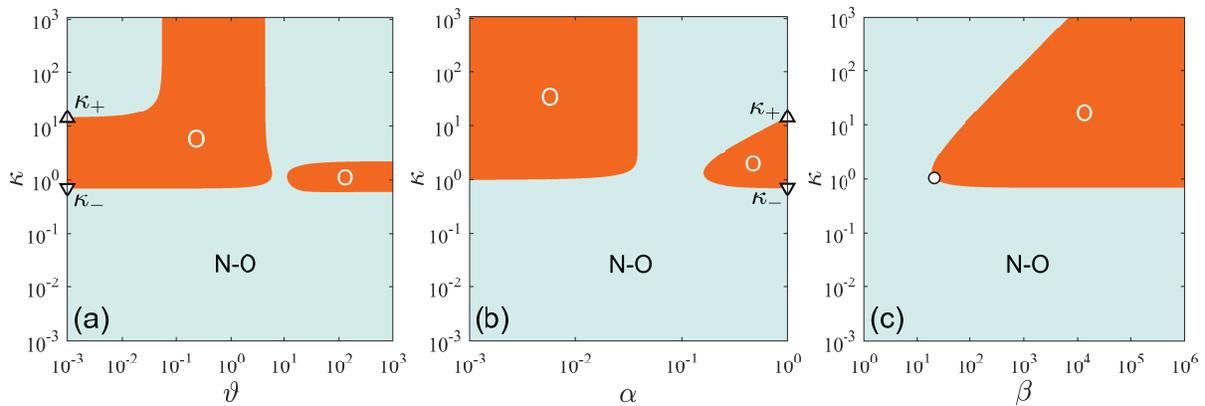}
 \caption{ Diagrams of the different regimes of the dynamical structure of the active suspension in a Maxwell-type viscoelastic fluid that can be probed depending on the values of (a) $\vartheta$ and $\kappa$ for $\beta = 10$ and $\alpha=0.1$; (b) $\alpha$ and $\kappa$ for $\beta = 10$ and $\vartheta= 10$; and (c) $\beta$ and $\kappa$ for $\vartheta = 10$ and $\alpha=0.1$. The areas marked by O correspond to the values of the parameters at which oscillations in the intermediate scattering function emerge, whereas no oscillations occur at those marked by N-O. In \ref{fig:3}(a) and \ref{fig:3}(b), the symbols $\triangle$ and $\bigtriangledown$ correspond the the characteristic wave numbers $\kappa_+ = 14.1244$ and $\kappa_- = 0.708$ given by Eq. (\ref{eq:kappa}) for a Newtonian fluid with $\beta = 10$. In \ref{fig:3}(c), the symbol $\circ$ depicts the onset of oscillatory behavior.}
\label{fig:4}
\end{figure*}

We now analyze the influence of the viscoelastic parameter $\alpha$ on the dynamic structure of active suspension. To this end, in Fig. \ref{fig:3}(c) we plot $\hat{S}(\bm{\kappa},\Omega)$ as a function of $\Omega$ for $\beta = 10$, $\vartheta = 10$ at $\kappa = 1$ and distinct values of $\alpha$ spanning the full interval $0 \le \alpha \le 1$. The reference curve for $\alpha = 1$ (Newtonian solvent of viscosity $\eta_0 = \eta_{\infty}$), which is depicted as a solid thin line in Fig. \ref{fig:3}(c), exhibits the characteristic shape with a low-frequency plateau $\hat{S}(\bm{\kappa},\Omega \rightarrow 0) = S_0 \kappa^{-2} \beta (1+\beta)^{-1}$, followed by a peak around $\Omega \approx 1$ and a high-frequency diffusive decay $\propto \Omega^{-2}$. With decreasing values of $\alpha$, we find systematic deviations from the Newtonian case, which result in the complete disappearance of the peak for $\alpha = 0.1$, and its re-emergence for $0 \le  \alpha < 10^{-2}$ but with a height and a frequency that are both smaller than those of the active suspension in a Newtonian solvent, see inset of Fig. \ref{fig:3}(c). This translates into an increase of the oscillation period of the intermediate scattering function for $\alpha < 0.01$, as well as a decrease in the amplitude of the oscillations, as shown in Fig. \ref{fig:3}(d). In addition, for $\alpha < 0.1$ we observe the appearance of a second plateau in the dynamic structure factor within the frequency interval $\alpha^{-1/2}  \vartheta^{-1} \lesssim \Omega \lesssim \alpha^{-1} \vartheta^{-1}$, which becomes increasingly broader with decreasing values of $\alpha$. Furthermore, for $\Omega \gtrsim \alpha^{-1} \vartheta^{-1}$, we find the high-frequency diffusive behavior $\hat{S}(\bm{\kappa},\Omega) \approx S_0 \alpha^{-1} \beta^{-1} \kappa^2 \Omega^{-2}$. This results in the complex shape of the intermediate scattering function shown in Fig. \ref{fig:3}(d) for $\alpha < 10^{-1}$, where a double temporal decay is observed at time-scales that are shorter than the corresponding oscillation period. Note that for the specific value $\alpha = 0$, which corresponds to the more widely known Maxwell model with a single viscosity $\eta_0 >0$ and $\eta_{\infty} = 0$, a marked change in the behavior of $\hat{S}(\bm{\kappa},\Omega)$ happens, where the second plateau at high frequencies completely disappears. This in turn leads to a pronounced change in the short-time behavior of $f(\bm{\kappa},\tau)$ with respect to the curves at non-zero $\alpha$, where a single decay instead of two occurs and the value to which it converges as $\tau \rightarrow 0$ is $f(\bm{\kappa},\tau \rightarrow 0) \rightarrow \frac{1}{2}\sqrt{\frac{\pi}{2}}$, i.e. twice smaller than the value for $\alpha > 0$.

The previous results evidence that the viscoelasticity of the solvent, characterized by the specific values of the parameters $\vartheta$ and $\alpha$, can strongly impact in an intricate manner the dynamical structure of the active suspension with respect to that in a Newtonian solvent. To illustrate this, in Fig. \ref{fig:4} we plot three diagrams that represent the possible regimes that can be probed in the active suspension by means of the intermediate scattering function over several orders of magnitude of the parameters $\vartheta$, $\alpha$, $\beta$ and $\kappa$. First, by varying the parameter $\vartheta$ of the viscoelastic fluid at constant $\beta > 1$ and $\alpha < 1$, two disconnected regions where oscillations in $f(\bm{\kappa},\tau)$ arise can be identified in the $\vartheta-\kappa$ plane, as shown in Fig. \ref{fig:4}(a) for $\beta = 10$ and $\alpha =0.1$. Out of these two regions, no oscillations in $f(\bm{\kappa},\tau)$ happen, and therefore for such possible values of the parameters $\vartheta$ and $\kappa$ the the active suspension effectively behaves as a passive system even when the active diffusion coefficient is bigger than the thermal one. Moreover, by varying $\alpha$ at constant $\vartheta > 0$, we also find two disconnected regions in the $\alpha-\kappa$ plane where oscillatory behavior of the intermediate scattering function takes place, see Fig. \ref{fig:4}(b) for $\beta = 10$ and $\vartheta = 10$. Once again, we find that out  of these two regions in the parameter space, the oscillations are suppressed even when $\beta > 1$. Finally, in Fig. \ref{fig:4}(c) we demonstrate that in a viscoelastic fluid, even though the shape of the the diagram is qualitatively similar to that of a Newtonian case (see Fig. \ref{fig:2}(e)), the numerical values of $\beta$ and $\kappa$ at which oscillations occur are strongly modified by the specific values of $\alpha$ and $\vartheta$. In particular, the onset of the oscillations, which critically occurs at $(\beta = 1, \kappa = 1)$ in a Newtonian solvent, is shifted to $(\beta \approx 20, \kappa = 1)$ in a Maxwell viscoelastic fluid with $\alpha = 0.1$ and $\vartheta = 10$. This implies that a higher particle activity is required in a viscoelastic solvent than in a Newtonian one to distinguish the dynamical structure of the active suspension from that of an inert suspension of passive Brownian particles under similar conditions.

\section{Concluding remarks}\label{sect:summ}

In this paper, we have analyzed 
spatio-temporal properties of  the number density fluctuations of a dilute suspension of active particles in a viscoelastic solvent. The proposed model for the particle motion captures the influence of rotational diffusion on the self-propulsion of the active particles as well as temporal memory effects on the friction exerted by the surrounding viscoelastic medium. In the case of a Maxwell-type model for the viscoelastic fluid, we find that, unlike a passive suspension in a Newtonian solvent, the dynamic structure factor and the intermediate scattering function have nontrivial dependencies on the parameters that characterize the viscoelasticity of the solvent and the activity of the particles. This approach allows us to identify regions in the parameter space of the suspension where the intermediate scattering function exhibits either a temporal oscillatory behavior or a monotonic decay in time. While the former is a hallmark of the particle activity, the latter can be regarded as an effective passive regime. Therefore, our results could serve as a reference to probe the non-equilibrium nature of active suspensions in non-Newtonian fluids by means of, \emph{e.g.}, DLS and DDM experiments. We point out that Maxwell-like models have been successful in providing a stochastic description for fluctuations of hydrodynamic variables in various complex fluids under external gradients, \cite{brand1987,camacho2005,rodriguez2006}. Thus, our approach could also be useful to investigate the effects of external gradients in the dynamics of active suspensions, \emph{e.g.}, particle concentration gradients, hydrodynamic flow, or  temperature gradients, which are of great relevance to active matter systems. Another aspects that could be investigated in future work are the effect of long-ranged correlations in the propulsion velocity of the particles \cite{sevilla2019,gomezsolano2020}, negative friction memory \cite{mitterwallner2020} as well as spatially dependent friction \cite{breoni2021}, on the dynamic structure factor of active suspensions in complex fluids.

\section*{Acknowledgments}
J. R. G.-S. acknowledges support from DGAPA-UNAM PAPIIT Grant No. IA104922.

\end{document}